# Big Data Analytics in Bioinformatics: A Machine Learning Perspective

Hirak Kashyap, Hasin Afzal Ahmed, Nazrul Hoque, Swarup Roy, and Dhruba Kumar Bhattacharyya

**Abstract**—Bioinformatics research is characterized by voluminous and incremental datasets and complex data analytics methods. The machine learning methods used in bioinformatics are iterative and parallel. These methods can be scaled to handle big data using the distributed and parallel computing technologies.

Usually big data tools perform computation in batch-mode and are not optimized for iterative processing and high data dependency among operations. In the recent years, parallel, incremental, and multi-view machine learning algorithms have been proposed. Similarly, graph-based architectures and in-memory big data tools have been developed to minimize I/O cost and optimize iterative processing.

However, there lack standard big data architectures and tools for many important bioinformatics problems, such as fast construction of co-expression and regulatory networks and salient module identification, detection of complexes over growing protein-protein interaction data, fast analysis of massive DNA, RNA, and protein sequence data, and fast querying on incremental and heterogeneous disease networks. This paper addresses the issues and challenges posed by several big data problems in bioinformatics, and gives an overview of the state of the art and the future research opportunities.

✦

## 1 INTRODUCTION

As we enter into the information age, data are being generated by variety of sources other than people and servers, such as sensors embedded into phones and wearable devices, video surveillance cameras, MRI scanners, and set-top boxes. Considering the annual growth of data generation, the digital universe - data we generate annually - will reach 44 zettabytes, or 44 trillion gigabytes by the year 2020, which is ten times the size of the digital universe in 2013 [2]. The fast transition into the information age has been fueled by the digitization of all of our devices and communication technology. Yesteryears technologies, such as analog telephony and film cameras, have been digitized. The advent of the Internet, followed by the WWW boom digitized our mailing systems, televisions, banking systems, and retailing, leading to storage and transmission of voluminous data. High performance technologies are used in scientific research, such as fast data capturing tools and very high resolution satellite data recording.

Apart from digitization of services and enterprises, a new trend has emerged recently to network all the man-made things around us, such as cars, home appliances, weapons, traffic lights, and power meters. These things communicate with each other to share data captured through various sensors, in order to take intelligent operational decisions by themselves. This network has been termed as the Internet of Things (IoT) [3]. The first networked appliance, a coke vending machine, was deployed at the Computer Science department of Carnegie Mellon University in the year 1990[1]. The IoT is growing fast and machine-to-machine connections will reach 1.2 billion in 2017, up from only 200 million in 2012 [4].

However, it should be noted that not all data, that we generate, are useful for descriptive or predictive analysis. Only a part of the data in the digital universe is useful, when tagged, termed as target-rich data. Metadata are more target-rich than the data itself. According to Turner et al. [2], approximately all of the target rich data were general IT data in the year 2014; however, by the year 2020, IoT data will occupy more than 20% of the target-rich data lake. Figure 1 shows the forecast made in the IDC report [2] regarding the size of the digital universe and the target-rich portion of it by the year 2020.

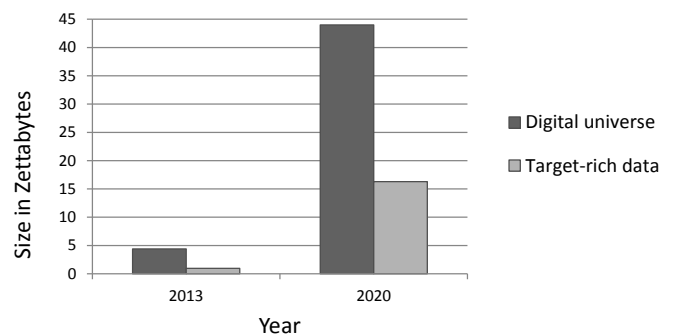

Fig. 1. Size of the digital universe in 2013 and 2020

The significance of these data are paramount as they embed the real life scenarios, such as environmental changes, cyber attacks, consumer drifts, and forthcoming epidemics,

- H. Kashyap, H. A. Ahmed, N Hoque, and D. K. Bhattacharyya are with Department of Computer Science and Engineering, Tezpur University, India - 784028.
  E-mail: {hirak, hasin, nhoq, dkb}@tezu.ernet.in
- S. Roy is with Department of Information Technology, North Eastern Hill University, Shillong-22, India.
  E-mail: swarup@nehu.ac.in

1. www.cs.cmu.edu/ coke/

and also because they are being generated and shared in real time. Consequently, these data are being heavily used for decision making and intelligent control.

Due to this high availability of information intensive data stream and the advances in high performance computing technologies, big data analytics have emerged to perform real time descriptive and predictive analysis on massive amount of data, in order to formulate intelligent informed decisions. Big data refers to a high volume of heterogeneous data formed by continuous or discontinuous information stream. In the literature, big data has been characterized as either 3Vs or 4Vs [5], [6]. The 3Vs refer to Volume, Velocity, and Variety; whereas the 4th V in the later definition refers to Veracity, i.e., reliability of the accumulated data. Additionally, there are two very important characteristics of big data that are not covered by this traditional definition. First, big data are incremental, i.e., new data are dynamically added to the big data lake from time to time. Second, big data are geographically distributed. These characteristics separate big data from traditional databases or data-warehouses.

## 1.1 Big data in bioinformatics

The volume of data is growing fast in bioinformatics research. Big data sources are no longer limited to particle physics experiments or search-engine logs and indexes. With digitization of all processes and availability of high throughput devices at lower costs, data volume is rising everywhere, including in bioinformatics research. For instance, the size of a single sequenced human genome is approximately 200 gigabytes [7]. This trend in rising data volume is also supported by decreasing computing cost and increasing analytics throughput with growing big data technologies. Biologists no longer use traditional laboratories to discover a novel biomarker for a disease, rather they rely on huge and continuously growing genomic data made available by various research groups. Technologies for capturing bio data are becoming cheaper and more effective, such as automated genome sequencers, giving rise to this new era of big data in bioinformatics.

The data size in bioinformatics is increasing dramatically in the recent years. The European Bioinformatics Institute (EBI), one of the largest biology-data repositories, had approximately 40 petabytes of data about genes, proteins, and small molecules in 2014, in comparsion to 18 petabytes in 2013 [8]. Their total storage size is doubling every year. Figure 2 shows the increasing trend in their genome and expression data store.

EBI has installed a cluster, the Hinxton data centre cluster, with 17,000 cores and 74 terabytes of RAM, to process their data. Its computing power is increased in almost every month. More importantly, EBI is not the only organization involved in massive bio-data store. There are many other organizations, who are storing and processing huge collections of biological databases and distributing them around the world, such as National Center for Biotechnology Information (NCBI), USA and National Institute of Genetics, Japan.

Availability of high volume of data is helpful for more accurate analytics, particularly in a highly sensitive field of

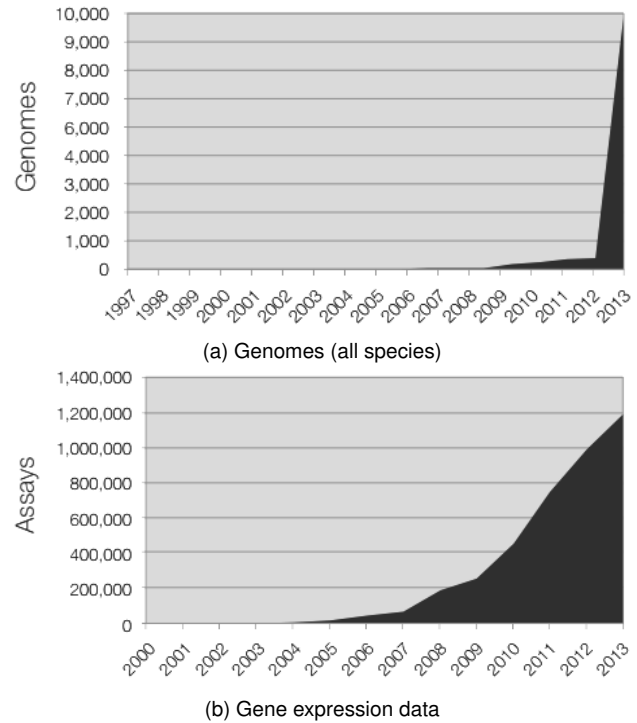

(a) Genomes (all species)

(b) Gene expression data

Fig. 2. Quantity of data stored by EBI over the years [8]

research like bioinformatics. However, the big data challenges here are much different from other well known big data problems, such as particle physics data captured at CERN or high resolution satellite data received at NRSC/ISRO open data archive[2]. The difference comes mainly in two aspects. First, bioinformatics data are highly heterogeneous in nature. Many analytics problems in bioinformatics require multiple heterogeneous and independent databases for inference and validation. Moreover, bioinformatics data are generated by many uncontrolled organizations and consequently, the same types of data are represented in different forms by their sources. Second, bioinformatics data, massive and growing in terms of dimension and number of instances, is geographically distributed all over the world. While part of these data may be transferred over the Internet, the remaining are not transferable due to their size (and hence inefficient), cost, privacy, and other ethical issues [9]. This sometimes forces to perform part of the analysis remotely and share the results. Therefore, big data problems in bioinformatics are not only characterized by volume, velocity, and variety, but also by geographically distributed data.

In order to tackle these challenges of big data in bioinformatics, cloud computing technologies have been used, with a lot of success. The best policy is to use cloud for both data store as well as for computation [9]. In fact, this policy helps to handle the big data challenges imposed by bioinformatics research over massive, growing and remotely distributed data. BGI, formerly known as Beijing Genomics Institute, one of the world's premiere genome sequencing centers, has installed a cloud-based analysis workflow called Gaea, using Hadoop framework. Gaea can

2. bhuvan.nrsc.gov.in

be used to perform large-scale genome analysis in parallel across hundreds of cloud-based computers. Another notable cloud-based genome analytics solution is provided by Bina Technologies[3], a Stanford University and UC Berkeley spin-off, in terms of a hardware component, called Bina box, to do the pre-processing on genome data and a cloud-based component to perform analytics on the pre-processed data. Bina box also reduces the size of genome data for their efficient transfer to the cloud component. This solution claims to improve the throughput of genome analytics by orders of magnitude higher than the traditional approaches [10].

## 1.2 Types of big data in bioinformatics

There are primarily five types of data that are massive in size and used heavily in bioinformatics research: i) gene expression data, ii) DNA, RNA, and protein sequence data, iii) protein-protein interaction (PPI) data, iv) pathway data, and v) gene ontology (GO). Although, other types of data such as human disease network and disease gene association network are also used, and highly important for many research directions including disease diagnosis.

In gene expression analysis, the expression levels of thousands of genes are analyzed over different conditions, such as separate developmental stages of treatments or diseases. Microarray-based gene expression profiling is usually used to record the expression levels for analysis. There are three types of microarray data, namely gene-sample, gene-time, and gene-sample-time. Gene expression profiles over sample space record the expression levels for varying external conditions, whereas over time space, they record the expression levels at different instances of time. Gene expression analysis can identify genes that are affected from pathogens or viruses, by comparing the expression values from infected and uninfected cells. The analysis results may be used to suggest biomarkers for disease diagnosis and prevention, among others. There are many public sources for microrarray databases, such as ArrayExpress[4] from EBI, Gene Expression Omnibus[5] from NCBI, and Stanford Microarray Database[6].

In sequence analysis, DNA, RNA or peptide sequences are processed using various analytical methods to understand their features, functions, structures, and evolution. DNA sequencing is used in study of genomes and proteins and their associations with diseases and phenotypes and identification of potential drugs, evolutionary biology, identification of micro species present in a sample environment, forensic identification, etc. Sequence analysis methodologies include sequence alignment and biological database search, among others. Although RNA sequencing is mainly used as an alternative for microarrays, it can be used for additional purposes also, such as mutation identification, identification of post-transcriptional mechanisms, detection of viruses and exogenous RNAs, and identification of Polyadenylation. Sequence analysis is more effective than microarray analysis, since sequence data embed richer information. However, it requires more sophisticated analytic tools and computing infrastructures, in order to deal with massive amount of sequence data [11]. Important sequence databases include DNA Data Bank of Japan[7], RDP[8], and miRBase[9].

PPIs provide crucial information regarding all biological processes. Therefore, forming and analyzing PPI networks can give a proper understanding of protein functions. PPIs are intrinsic to the interactomics of living cell. Therefore, anomalous PPIs are the basis of various diseases, such as Alzheimer's disease and cancer. PPIs have been studied in different fields of research, such as bioinformatics, biochemistry, quantum chemistry, and molecular dynamics, thus giving rise to high volume of heterogeneous data regarding the interactions. Important PPI repositories are DIP[10], STRING[11], and BioGRID[12], among others

Pathway analysis is useful for understanding molecular basis of a disease. Additionally, pathway analysis identifies genes and proteins associated with the etiology of a disease, predicts drug targets, and helps to conduct targeted literature searches. Further, it helps to integrate diverse biological information and assign functions to genes. The most notable pathway data sources are KEGG [12], Reactome [13], and Pathway Commons [14].

The GO database[13] provides dynamic, structured, and species-independent gene ontologies for associated biological processes, cellular components, and molecular functions. The GO database uses controlled vocabularies to facilitate querying at different levels. A vast number of tools uses the GO database for bioinformatics research. Most of these tools are third-party based, however the GO project itself maintains certain tools, such as AmiGO, DAG-Edit, and OBO-Edit. The GO-based tool chain is so huge that there exist tools, such as SerbGO [15], to search the appropriate GO tools for a particular bioinformatics problem. The GO database has highly been used for various purposes, such as to build ontologies for anatomies, to validate semi-supervised and unsupervised analytics results from data, and to develop timelines for model organisms, human diseases and plant growth environments.

## 1.3 Big data problems in bioinformatics

The solutions for cloud-based large-scale big data analytics, such as Bina box for genome analysis, are very recent. There are several other big data problems in the domain of bioinformatics that are yet to be explored. Considering the recent big data boom in bioinformatics, as discussed above, there is an urgent need to address many of these problems. In this paper, we categorize the big data analytics problems in bioinformatics into seven categories. They are discussed below.

### 1.3.1 Microarray data analysis

The size and number of microarray datasets are growing rapidly, mainly due to decreasing cost and widespread use of microarray experiments. Moreover, microarray experiments are also been performed for gene-sample-time space, in order to capture the changes in expression values over time or over different stages of a disease. Big data technologies are required for fast construction of co-expression and regulatory networks using voluminous microarray.

---

3. www.bina.com
4. www.ebi.ac.uk/arrayexpress
5. www.ncbi.nlm.nih.gov/geo
6. smd.princeton.edu
7. www.ddbj.nig.ac.jp
8. rdp.cme.msu.edu
9. www.mirbase.org
10. dip.doe-mbi.ucla.edu
11. string.embl.de
12. thebiogrid.org
13. www.geneontology.org

As gene expression data are being captured at different progression stages of a disease over time, there has been an opportunity to identify the genes that are affected by the disease, in order to identify biomarkers for the disease. Computationally, the addition of the third dimension, time, makes the analytics much higher in complexity than the traditional analysis of gene complexes.

### 1.3.2 Gene-gene network analysis

Gene regulatory networks (GRN) alterations underlie many anomalous conditions, such as cancer. Inferring GRN and their alterations from high-throughput microarray data is a fundamental but challenging task. With the rapid growth of high throughput sequencing technologies, system biologist are now able to infer gigabytes of data. In many cases, movement of such large volume of data is not feasible. Integration of large multiple GRNs from different sources help in reconstruction of a unified GRN. Reconstruction of GRNs locally and then their integration through cloud infrastructure may help system biologists to better analyze a diseased network.

Additionally, the inference can be translated to genomic medicine. Although there exist many GRN inference mechanisms, their relative strength are unknown, due to the lack of large-scale validation. To find the most effective inference mechanism to identify the abnormal networks and to prioritize the target proteins for druggabilty are demanding issues and need to be addressed using fast, reliable, and scalable architectures.

Gene co-expression network analysis estimates the correlation among different gene-gene networks obtained from gene-expression analysis. Differential co-expression analysis finds the changes incurred by the gene complexes over time or over different stages of a disease. This helps in finding the relations between gene complexes and traits of interest. Gene complexes of different species can also be studied to find genotypic similarities. Gene co-expression network analysis is a complex and highly iterative problem and requires large-scale data analytics systems.

### 1.3.3 PPI data analysis

PPI complexes and changes in them inhibit high information content about various diseases. PPI networks are being studied in various domains of life sciences with production of voluminous data. The volume, velocity, and variety of data make PPI complex analytics is a genuine big data problem. It demands for an efficient and scalable architecture to provide fast and accurate PPI complex generation, validation, and rank-aggregation.

### 1.3.4 Sequence analysis

With the increasing volume (in order of petabytes) of DNA data deluge originated from thousands of sources, the present DNA sequencing tools have been found inadequate.

So, development of a high throughput and compact architecture for DNA sequence analysis with renewed focus for big data management is a bioinformatics problem with high demand in the recent days.

RNA sequencing technology has emerged as a strong successor to the microarray technology, due to its more accurate and quantitative gene expression measurements. However, RNA sequence data also contain additional information, which are often overlooked, and require complex machine learning techniques to be extracted. Big data technologies can be used to identify mutations, allele-specific expressions, and exogenous RNA contents, such as viruses, from RNA sequence data using sophisticated machine learning methods.

The next generation genome sequencing provides information on the complete genome of an individual, in orders of magnitude bigger in size than microarray based methods for genetic assessment. Large scale methods are needed to study the specific changes in genome sequences due to a particular disease and to compare with the existing results of the same or different related diseases.

### 1.3.5 Evolutionary research

The recent advances in molecular biological technologies have become a prominent source for the generation of big data. Huge amount data has been generated by various projects at microbial level, such as whole genome sequencing, microarrays, and metabolomics. Bioinformatics has emerged as a significant platform for analysis and archival of this wealth of information. An important big data problem in bioinformatics has been the study of functional trends of adaptation and evolution using microbial research, by investigating primitive organisms.

### 1.3.6 Pathway analysis

Pathway analysis associates genetic products with phenotypes of interest, in order to predict gene function, identify biomarkers and traits, and classify patients and samples. The genetic, genomic, metabolomics, and proteomic data has increased rapidly and big data technologies are required to perform association analysis on huge volumes of these data.

### 1.3.7 Disease network analysis

Large disease networks have been formulated for many species, including human. These networks are continuously growing and new networks are being added by different sources in their own format. The multi-objective associations among diseases in heterogeneous networks are useful for understanding the relations among diseases across networks. Traditional network analytics techniques would not perform well over unstructured and heterogeneous data without compromising information quality, and intelligent and efficient analytics are required. Big data technologies are required to effectively deep mine the associations among heterogeneous disease networks.

Complex networks of molecular phenotypes characterize causal or predictive genes or mechanisms for disease-associated traits. Ability for fast processing of these data allows researchers to analyze more datasets, that were not possible to analyze before. Although large collections of these data can be analyzed with existing technologies, techniques for data integration are still inefficient. Optimal integration methods are required to analyze multiple, heterogeneous omics databases.

In addition to that, new high-throughput methods collect personalized phenotypes of huge number of individuals. Large scale machine learning tools are needed to recognize and visualize complex data patterns for the purpose of disease genesis analysis and diagnosis.

Although some of these bioinformatics problems existed before the big data era, their complexity and efficiency have significantly scaled up with the emerge of big data. On the other hand, the other problems have been made possible by the availability of massive amount of data. In either case, sophisticated big data analytics technologies are of urgent need to handle these large scale problems.

## 1.4 Techniques for big data Analytics

Supervised, unsupervised, and hybrid machine learning approaches are the most widely used tools for descriptive and predictive analytics on big data. Apart from that, various techniques from mathematics have been used in big data analytics. The problem of big data volume can be somewhat minimized by dimensionality reduction. Linear mapping methods, such as principal component analysis (PCA) and singular value decomposition, as well as non-linear mapping methods, such as Sammon's mapping, kernel principal component analysis, and laplacian eigenmaps, have been widely used for dimensionality reduction.

Another important tool used in big data analytics is mathematical optimization. Subfields of optimization, such as constraint satisfaction programming, dynamic programming, and heuristics & metaheuristics are widely used in AI and machine learning problems. Other important optimization methods include multi-objective and multi-modal optimization methods, such as pareto optimization [16] and evolutionary algorithms [17], respectively.

Statistics is considered as a counterpart to machine learning; differentiated by data model versus algorithmic model respectively. The two fields have subsumed ideas from each other. Statistical concepts, such as expectation-maximization and PCA, are widely adopted in machine learning problems. Similarly, machine learning techniques, such as probably approximately correct learning are used in applied statistics. However, both of these tools have been heavily used for big data analytics.

Big data analytics has a close proximity to data mining approaches. Mining big data is more challenging than traditional data mining due to massive data volume. The common practice is to extend the existing data mining algorithms to cope with massive datasets, by executing on samples of big data and then merging the sample results. This kind of clustering algorithms include CLARA (Clustering LARge Applications) [18] and BIRCH (Balanced Iterative Reducing using Cluster Hierarchies) [19]. Researchers have also emphasized on the reduction of computational complexity of data mining algorithms. For example, spectral regression discriminant analysis significantly reduces the time and space complexity by simplifying discriminant analysis to a set of regularized least squares problems [20]. Similarly, Shi et al. [21] reduce the space complexity of non-linear discriminant analysis from $O(n^2)$ to $O(n)$, to minimize computation and storage problem on large-scale datasets.

Nevertheless, time and space complexity of most of the machine learning and statistical methods are very high to be effective for real time analysis on large-scale dataset. In the recent years, distributed and parallel computing technologies have emerged as the prime solution to large-scale computing problems, due to their scalability, performance, and reliability. Therefore, efforts have been made to perform big data analytics using distributed computing, under strict performance and reliability constraints. Consequently, distributed data analytics algorithms have been proposed in the literature. Mining of distributed data in itself has emerged as a new paradigm of data analytics. It should be noted that, to be effective, the nodes should perform the computations independently, i.e., without constantly sharing intermediate data with peer nodes. Park and Kargupta [22] discuss the distributed algorithms for classifier learning, association rule mining, and clustering. Rana et al. propose a component-based system, designated as PaDDMAS, for developing distributed data mining applications [23]. Similar systems for distributed machine learning methods are proposed, such as MLbase [24]. Further, cloud computing infrastructure-based systems are also proposed for performing distributed machine learning, such as the Distributed GraphLab [25] framework that emphasizes on consistency and fault-tolerance in distributed analytics.

The main driving force for big data analytics has been the industry researches for massive-scale commercial applications. Although cluster and grid computing have existed for long, they are designed specifically for particular applications and require high cost and expertise. Therefore, the technologies for big data analytics did not evolve significantly in that period. When cloud computing infrastructure and distributed processing platforms, such as MapReduce [26], and their open source implementations became widely available, the research on big data analytics escalated. Iterative graph processing systems, for solving large scale practical computing problems, have also been proposed. The proprietary graph processing architecture developed at Google, known as Pregel [27], addresses distributed processing of large scale real-life graphs. An open-source counterpart of Pregel is Apache Giraph[14], which provides additional features, such as edge oriented input and out-of-core computation.

Moreover, the rising data volume has contributed to the increasing demand for big data analytics. In the recent years, distributed file system technologies, such as HDFS [28] and QFS [29], as well as NoSQL databases for unstructured data, such as MongoDB[15] and CouchDB[16] have been widely used for big data analytics.

Machine learning libraries have been developed for big data analytics. The most notable machine learning library for big data analytics is Apache Mahout [30], which contain implementations of various machine learning techniques, such as classifiers, clustering, and recommender systems, which are scalable to large scale datasets. MLlib[17] is a similar library to perform machine learning on big data on the Apache Spark platform, a MapReduce variant for iterative and fast computations on big data. However, these libraries still lack many important machine learning methods and more contributions are needed from the community.

---

14. giraph.apache.org
15. www.mongodb.org
16. couchdb.apache.org

### 1.5 Contributions

This paper provides an in depth study on the sources and types of big data in bioinformatics, the existing machine learning and big data techniques to analyze them, and the limitations and future research. The contributions of this paper are listed below.

1) Two additional characteristics to the traditional definition of big data are introduced. Accordingly, big data are also incremental and geographically distributed.
2) The problems in bioinformatics, that face the challenges of huge, ever growing, and heterogeneous datasets, are categorized into seven classes. Research issues in each of these problem categories are identified.
3) State of the art big data technologies are classified into three classes based on their overall system architectures. The generic architectures for each of the classes are introduced.
4) Machine learning methods for large scale data analytics are presented. The limitations of the traditional methods and their incremental versions for fast, scalable, and accurate big data solutions are discussed.
5) Big data tools and machine learning techniques available for each category of the bioinformatics problems and the scope for future contributions are discussed.

### 1.6 Organization of the paper

The paper is organized as follows. In Section 2, generic architectures for the existing big data analytics computational models are presented. The traditional as well as the big data oriented machine learning methods, along with discussion on their capabilities and limitations are discussed in Section 3. In Section 4, the issues and challenges associated with big data analytics are discussed. Existing big data tools for bioinformatics are presented in Section 5 and our conclusions on the study are presented in Section 6.

## 2 ARCHITECTURES FOR BIG DATA ANALYTICS

Big data analytics systems have been proposed with several architectures. However, many of them share common computational models. Based on our study, we classify big data solutions into three major architectures. Each of them has their own advantages as well as limitations, and their suitability depends on the nature and requirements of the algorithm to be implemented. These are discussed below.

17. spark.apache.org/mllib

### 2.1 MapReduce architecture

MapReduce is a data-parallel architecture, originally developed by Google [31]. Parallelism is achieved by multiple machines or nodes performing the same task on different data. Apache Hadoop[18] is a highly used open-source implementation of MapReduce. A MapReduce daemon runs on the nodes all the time. There is one master node that performs the configuration and control throughout the execution of the problem. The other nodes are called the worker nodes and perform actual computation on data. The master node also splits the data, assigns them to worker nodes, and puts them into the global memory as (key, value) pairs. Figure 3 depicts the basic architecture of MapReduce, where Wi's are the worker nodes.

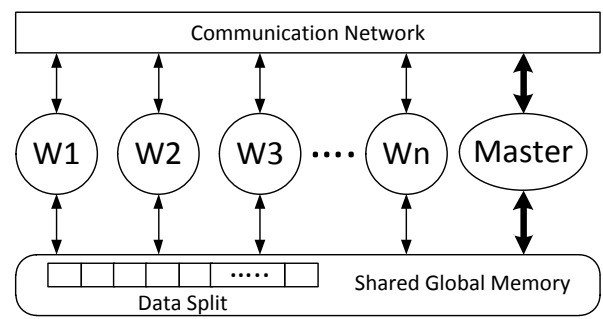

Fig. 3. MapReduce architecture

MapReduce works in rounds, each consisting of two phases, namely map and reduce phases. A node can be used in both map and reduce phases. Each phase consists of three states: input, computation, and output. There is one synchronization barrier between any two consecutive phases. During synchronization, local memory of a node is cleared and written onto the global memory. The master node can read/write onto the global memory and communicate with the other nodes during all time. However, the worker nodes can read/write onto the global memory only during synchronization. In Figure 3, this has been distinguished using thick and thin arrows.

During the map phase, the problem is data-distributed among the worker nodes and the partial results generated by the worker nodes are stored in the global memory. During the reduce phase, the partial results are combined to obtain the overall result, to be stored in the global memory. If the intermediate results need to be further processed, the phases are repeated again.

MapReduce architecture performs well when the data size is huge and the problem in hand is embarrassingly parallel. The architecture provides fault-tolerance by re-doing the computation (done by the failing node) for the phase on another node. However, the architecture has limitations for problems involving high computational dependencies among data. Moreover, the architecture cannot be used to express iterative computations and becomes inefficient with high I/O overhead.

18. hadoop.apache.org

Efforts have been made to mitigate the limitations of the MapReduce architecture and improve its performance. Twister [32] optimizes iterative computations on the MapReduce architecture by using in-memory computations, rather than writing onto the distributed memory after each phase. However, Twister has fault-tolerant issues due to in-memory processing. Apache Spark[19] extends Hadoop by using Resilient Distributed Database (RDD) [33] to allow in-memory processing as well as fault-tolerance by reconstructing a faulty partition in case of node failure.

## 2.2 Fault tolerant graph architecture

While MapReduce and its different implementations process data in batch mode, they are not very expressive when complex computational dependencies exist among data. Most of the machine learning and statistical methods inhibit high data dependencies. Therefore, MapReduce is not the best architecture for them. Alternate architectures are needed to process the complex and iterative problems efficiently, while supporting fault tolerance. Fault tolerance is important for scalability also, since it allows to use unreliable networks, such as the Internet.

In order to achieve that, a fault tolerant graph-based architecture, called GraphLab, was first proposed by Low et al. [34] and later many other big data solutions adopted similar architectures. In this architecture, the computation is divided among nodes in a heterogeneous way, with each of them performing some particular tasks. The data model is divided into two parts, i) a graph with computing nodes and ii) a shared global memory (distributed). The generic architecture is depicted in Figure 4. The Ni's represent the computing nodes and the dotted arrows show the dependencies among the nodes, whereas actual communication is performed via the communication network.

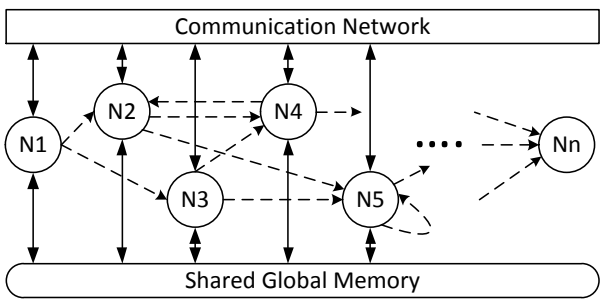

Fig. 4. Architecture for Graph with global shared memory

Similar to MapReduce, the computation is carried out in execution cycles in a synchronous manner. The shared database is initialized with the input data. At the beginning of each cycle, a node first reads the shared database and then performs computation using its own and its neighbor's data. Then the results are merged and then written back to the global shared database, for use in the next execution cycle. If a particular node fails in one cycle, it is recomputed and the dependent nodes lose one cycle. Although it reduces the

19. spark.apache.org

efficiency by a cycle, the fault tolerance is guaranteed. If a node fails permanently, then it is replaced.

This architecture provides high expressiveness for complex problems with data dependency and iterations. However, the architecture demands high disk I/O and therefore, it is not optimized for performance. To the best of our knowledge, an improvement using RDD to facilitate in-memory processing and fault tolerance is not yet proposed.

Apart from GraphLab, other major graph-based big data solutions are Pregel and Giraph. Graph packages are also developed for the MapReduce architecture, such as GraphX and the Hama[20] graph package called Angrapa.

## 2.3 Streaming graph architecture

The graph-based architecture discussed above allows scalable distributed computation, complex data dependency among operations, efficient iterative processing, and fault tolerance. However, due to its high disk read/write overhead, it is not efficient for stream data. Although there are packages to perform analytics on stream data on the MapReduce architecture, such as Spark Streaming[21], they internally convert stream data to batches for processing. Stream applications require in-memory processing for high bandwidth. The well known Message Passing Interface (MPI) [35] is a good fit for this problem. At the application level, MPI has similar API as MapReduce and almost all MapReduce programs can also be implemented using MPI. Figure 5 depicts the graph-based architecture for large scale distributed processing, for high bandwidth and iterative applications with high data dependency among operations. This architecture is in line with the ever increasing computing speed and improved network bandwidth and reliability.

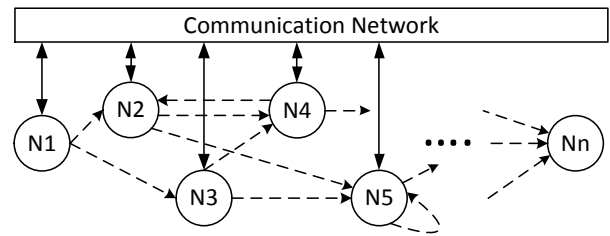

Fig. 5. Architecture for graph-based asynchronous processing

There are three major differences between this architecture and the previous one. First, in this architecture, a global shared memory is not used, rather the nodes exchange data using peer-to-peer communications directly. Second, the operations are performed in an asynchronous manner. The different data flows become synchronous only during their merge operations. Finally, in this architecture, data need not be stored into disks. As memories are becoming cheaper everyday, in-memory processing of large volume data is possible, which significantly increases the overall throughput.

20. hama.apache.org
21. spark.apache.org/streaming

The main disadvantage of this architecture is the absence of fault tolerance. If any one of the nodes fail, the process has to start from the beginning all over again. Consequently, this architecture is unsuitable in unreliable networks, such as the Internet. This in turn causes scalability issues. However, if a reliable network is available and the algorithm has high data dependency, then this architecture can provide higher throughput than the other architectures. This architecture can be implemented on standalone clusters using MPI to perform analytics on big data .

## 3 MACHINE LEARNING FOR BIG DATA ANALYTICS

Machine learning techniques have been found very effective and relevant to many real world applications in bioinformatics, network security, healthcare, banking and finance, and transportations. Over time, bioinformatics and health-related data are created and accumulated continuously, resulting in an incredible volume of data. Newer forms of big data, such as 3D imaging, genomics and biometric sensor readings are also fueling this exponential growth. Future applications of real-time data, such as early detection of infections/diseases and fast application of the appropriate treatments (not just broad-spectrum antibiotics) could reduce patient morbidity and mortality. Already, real-time streaming data monitors neonates in the ICU, catching life-threatening infections at real time. The ability to perform real-time analytics against such voluminous stream data across all specialties would revolutionize healthcare. Therein lies data with volume, velocity, and variety.

Machine learning is a field of computer science that studies the computational methods that learn from data [36]. There are mainly two types of learning methods in machine learning, viz., supervised and unsupervised learning methods [37]. In supervised learning, a method learns from a set of objects with class label, often called a training set. The acquired knowledge is used to assign label to unknown objects often called test objects. On the other hand, unsupervised learning methods do not depend on the availability of prior knowledge or training instances with class labels. All these machine learning methods require preprocessing of datasets for effective results. Feature selection is one of the important preprocessing tasks that leads to improved result and reduced time requirement. Hybrid learning methods, such as Deep learning, have become popular in the recent years and provide significantly high accuracy.

Advanced data capturing technologies have led to accumulation of a very high volume of data, growing rapidly over time. Although the computational technologies have improved over time, this improvement is not proportionate to the rate of increase in data volume. The traditional machine learning methods are found inadequate in handling voluminous data using the current computational resources [38]. Figure 6 depicts the contrast between traditional data mining and mining of big data.

To apply a traditional, or enhanced a new machine learning method to analyze big data, following properties are desirable.
- Scalable to high volume: The method should be able to handle large chunk of data with low space complexity and less disk overhead.

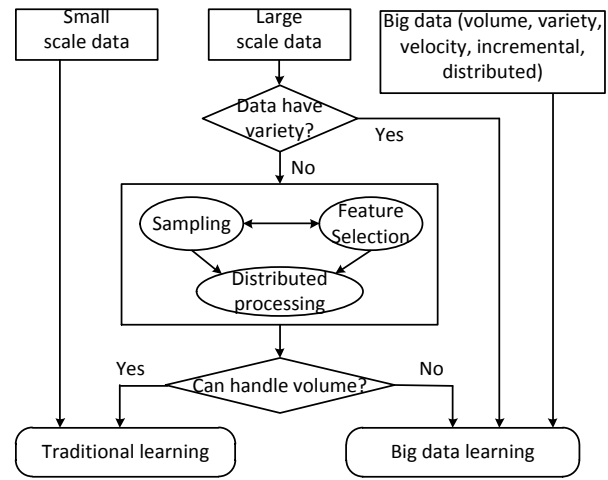

Fig. 6. Traditional data mining and mining of big data

- Robust with high velocity: The method should have low time complexity and be able to digest and process stream data in real time without any degradation in performance.
- Transparent to variety: Big data can be semi-structured or unstructured in nature. However, most traditional machine learning methods are able to process datasets with a fixed schema, which is normally generated from a single source. By the term schema, we refer to an ordered set of features and the relations among them. A machine learning method for big data analytics should be able to handle data from multiple sources with different schema.
- Incremental: Typically, machine learning methods operate on entire datasets at once without accounting for the situation where dataset dynamically grows over time. A machine learning method for big data analytics should consider the inconsistent arrival of data over time and should be able to handle such data with minimum cost, without compromising quality.
- Distributed: A machine learning method should allow distributed processing on partial data and merging of the partial results. With big data sources distributed around the world, all data may not be available at a single location for big data analytics.

### 3.1 Feature Selection

The main objective of feature selection is to select a subset of most relevant and non-redundant features that can increase the performance of a learning method. A feature selection method can improve the performance of prediction models by removing irrelevant and redundant features with alleviating the effect of the curse of dimensionality, enhancing the generalization performance, speeding up the learning process, and improving the model interpretability [39]. Due to wide application of computer networks and Internet, data over Internet communication as well as in many other online services must deal with large volume of data with volume, velocity, and variety. Moreover, in many business applications, handling big data is an essential requirement but taking instant decision reliably on big data is still an open research issue. Such big data pose great challenges for feature selection in terms of performance, scalability, robustness, universality, nonlinearity, and cost and implementation complexity.

A feature selection plays a major role in identifying the most important features from a ultrahigh dimensional big dataset. The selected feature set can be used for processing large volume of data to take instant decision in short period of time. Especially, in big data analytics, relevant features can be selected from large data using both supervised learning as well as unsupervised learning. Hence, ranking the features based on their relevance and selecting the most relevant features can vastly improve the generalization performance.

Feature selection is also considered very important for big data analytics due to its characteristics of semi-infinite programming (SIP) problem [40]. The SIP is an optimization problem that can be stated either it is associated with a finite number of variables and an infinite number of constraints, or an infinite number of variables and a finite number of constraints. To address the SIP problem, Tan et al. [41] propose an efficient feature selection algorithm works iteratively and selects a subset of features, and solves a sequence of multiple kernel learning (MKL) subproblems. Authors claim that the proposed method converges globally under mild condition and yields low biasness on feature selection.

In bioinformatics, protein sequence analysis and PPI analysis are complex problems in functional genomics. A feature vector exhibits protein sequences with distinguished characters and the feature vector plays a major role during analysis of protein sequence. However, a major problem of PPI dataset is that it contains huge number of enormous features which increase not only the complexity of analysis but reduce prediction accuracy. To overcome this problem, Bagyamathi et al. [42] propose a new feature selection method combining Improved harmony search algorithm with rough set theory to tackle the feature selection problem in big data.

Barbu et. al. [43] propose a novel feature selection method with annealing technique for big data learning. In this method they reduce the dimensionality of an instance from $M$ to $k$ using an annealing plan to decrease greediness and remove the most irrelevant variables to facilitate complex computation. They termed the feature selection problem as a constrained optimization problem defined as $\beta$=arg min $L(\beta)$, such that, $|\{j : \beta_j \neq 0\} \leq k|$, where $k$ is the number of relevant features. The algorithm is extremely suitable for big data computation due to its simplicity and ability to reduce the problem size throughout the iteration.

Incremental learning is useful to predict behavior of big data in terms of adaptiveness. An incremental learning method considers subset of features selected incrementally from samples of data over time. For efficient analysis of high volume of data with random velocity and multiple varieties, incremental feature selection method selects those features that can predict the behavior of data efficiently. Zeng et al. [44] propose an incremental feature selection method called FRSA-IFS-HIS (AD) using fuzzy-rough set theory on hybrid information systems. The method has been found effective compared to non-incremental fuzzy-rough set feature selection method applied on big data.

## 3.2 Supervised Learning

In supervised learning, labeled training examples are used to train the learning algorithm. The objective of a supervised learning model is to predict the class labels of test instances based on knowledge gained from the available training instances. Within supervised learning family we can further distinguish between classification models which focus on prediction of discrete (categorical) outputs or regression models which predict continuous outputs. Among large number of models reported in the literature linear and nonlinear density-based classifiers, decision trees, naive Bayes, support vector machines (SVMs), neural networks and K-nearest neighbour (KNN) are the most frequently used methods in many applications [45] [46] [47] [48].

In big data analytics, we need some advanced supervised approaches for parallel and distributed learning such as Multi-hyperplane Machine (MM) classification model [49], divide-and-conquer SVM [50], and neural network classifiers. Among these SVM is one of the most efficient and widely used supervised learning method and several modified SVM methods have been introduced for big data analytics. Nie et al. propose a modified SVM called New Primal SVM for big data classification [51]. The method uses a novel linear computational cost primal SVM solver using two loss functions called L1-norm and L2-norm in Augmented Lagrange Multipliers (ALM). Individual detection of patients with Parkinson disease using SVM analysis was proposed by Haller et. al. [52]. In this work, the authors adopt a complex methodology including a chain of tract-based spatial statistics (TBSS) preprocessing of DTI fractional anisotropy data, feature selection of the most discriminative voxels, and subsequent SVM classification. Experimental results establish the effectiveness of the proposed method and feasibility of performing SVM individual classification of DTI data in patient diagnosis, which may merit future prospective and larger scale follow-up studies. Giveki et. al. [53] propose a weighted SVM based on mutual information and modified cuckoo search for automatic detection of diabetics diagnosis. The method first applies principal component analysis for feature selection from diabetes dataset and estimates the best feature weights using mutual information. Afterwards, the method is applied to classify patients where modified cuckoo search is used to find the best value for $C$ and $\gamma$ parameters of the proposed methods.

Another SVM-based decision support system for heart disease classification with integer-coded genetic algorithm to select crucial features was proposed by Bhatia et. al. [54]. The method uses an integer-coded genetic algorithm to select an optimal subset of features from Cleveland heart disease database which maximizes the SVM classification accuracy with a reduced number of features used by the SVM classifier to classify heart disease. Son et al. [55] use SVM to classify heart failure patients.

Distributed decision tree is another significant effort to improve the performance of decision tree induction when processing on big data by parallelizing the induction process and by performing the induction process in distributed environment. Ye et. al. propose techniques to distribute and parallelize Gradient Boosted Decision Trees(GBDT) [56]. It is very straightforward to convert GBDT to MapReduced model and in this method a MapReduced-based GBDT was employed for horizontal data partitioning. According to the authors, due to the high communication overhead of HDFS [57], Hadoop is not suitable for this algorithm.

Calaway et al. [58] propose fast, scalable and distributable decision tree called rxDTree which can estimate decision trees efficiently on big data. This algorithm is widely used in classification and regression problems of big data. It computes histograms to create empirical distribution functions of the data and builds the decision tree in a breadth-first fashion. The algorithm can be executed in parallel settings, such as a multicore machine or a distributed (cluster or grid) environment.

For big data, an intelligent agent could provide hint on areas of data that might the users would be very useful. If the dataset has categories for different user classes as class labels, then the labels can be used to train a decision tree to classify unseen data. But, the training set will be much larger than usual and hence, the rule generation for decision tree is a complex and time consuming process. To handle this problem, Hall et al. [59] propose a modified decision tree learning that generates rules from a set of decision trees built in parallel on tractable size training dataset.

## 3.3 Unsupervised learning

Unsupervised learning do not use the class labels of the objects for learning [60]. Clustering is an unsupervised technique that attempts to group objects to optimize the criterion that states that distance among objects in the same cluster is minimized and distance among objects in different clusters is maximized [61]. A major issue in clustering is the computation of distance between a pair of objects. Various proximity measures have been used for this purpose, such as Euclidean, Cosine, and city block distance. In traditional clustering, all the features are used while computing the distance between a pair of objects. A cluster is a group of objects that are close to each other with respect to their mutual distance. In other words, they are similar in nature over the entire set of features. However, in a number of applications, especially where number of available features in a dataset is very large, researchers are interested in finding groups of objects that are similar over subset of the available features [62]. This requirement has led to the emergence of another variant of clustering called biclustering, where each bicluster is associated with a subset of features.

Clustering and biclustering analyses two dimensional data, where each feature corresponds to an attribute of the objects. Value of an object over a feature is some form of quantification of the concerned attribute. With advent of data capturing technologies, it has been possible to trace down dynamic nature of the object attributes by capturing values over multiple consecutive time instances. This arrangement leads to generation of three dimensional datasets. Another variant of clustering, called triclustering operates on such datasets to generate triclusters. A tricluster is a group of objects that are not only similar over a subset of features, but are also similar across a subset of time points [63]. Triclustering promotes grouping of objects, features and time points simultaneously.

### 3.3.1 Existing clustering methods

Numerous clustering methods have been proposed so far in the field of machine learning. These clustering methods are mainly classified into partitional clustering, hierarchical clustering, density-based clustering, graph theoretic clustering, soft computing-based clustering, and matrix operation based clustering [64]. Partitional clustering methods assign objects to one of the $k$ clusters, where $k$ is a user given parameter, to iteratively optimize a criterion function. K-means [65] assign objects to the nearest cluster centroid iteratively until there is no more assignment possible. Partitioning Around Medoids (PAM) [66] is another partitional clustering method that uses medoids instead of centroids. PAM is robust, but inefficient in handling large dataset due to its $O(n^2)$ complexity. CLARA [66] and CLARANS [67] are two popular partitional clustering methods that use sampling for large datasets. CLARA draws a sample of objects on which PAM is applied. CLARAN uses sampling during neighborhood search operation. Both CLARA and CLARANS attempt to handle large dataset.

Hierarchical clustering methods can be classified into agglomerative and divisive methods [68]. Agglomerative approaches operate in bottom up direction on a tree and starts with nodes with individual objects. These nodes are iteratively merged to reach the root of the tree. In divisive approach, root with all the nodes are iteratively splitted to finally reach the leaf nodes. BIRCH [19] is a popular agglomerative hierarchical clustering method that constructs clustering feature (CF) tree first, which is operated in a bottom up fashion to extract the clusters. CURE [69] is another popular hierarchical clustering method that starts with some scattered objects to form clusters. These clusters are then shrunk towards theirs centers. DIANA [66] is a divisive hierarchical clustering method that splits largest cluster iteratively to find splinter groups.

Density-based clustering methods find clusters characterized as dense areas and separated by low dense regions [70]. Density of a node is measured using neighbourhood analysis. DBSCAN [71] is a very popular density-based clustering method that starts from an initial object and includes objects from it's neighbourhood iteratively if they satisfy a user defined threshold to form a cluster. DENCLUE [72] is another popular density-based clustering method that uses kernel density function. A cluster is defined as a local maximum of the density function.

Graph theoretic clustering methods use properties and concepts of graph theory [73]. CLIQUE [68] a graph theoretical clustering method tries to locate maximally complete subgraphs in the connectivity graph derived from actual datasets. These subgraphs correspond to detected clusters. Chameleon [74] is another graph theoretic agglomerative hierarchical clustering method that uses k-nearest neighbor graph. Here, edges are iteratively deleted if connecting nodes are not included in the k-nearest neighbour sets of each other.

Soft computing-based clustering methods use soft computation tools, such as fuzzy set and neural network. Fuzzy c-means [75] a soft computing-based clustering method, is a crisp method, which allows objects to belong to more than one cluster with the constraint that the sum of membership of an object across all the clusters is equal to one. This method tries to find a crisp partition that minimizes a cost function. SOM [76] is another very popular soft computing based clustering method that projects high dimensional vectors to a two dimensional grid space. Iteratively these objects are moved to dense regions that correspond to the clusters.

There are a lot of biclustering algorithms proposed by researchers. Cheng and Church [62] propose a biclustering method that iteratively deletes and adds objects and features in a greedy manner. OPSM [77] detects order preserving sub-matrices in data matrix which correspond to clusters. BIMAX [78] biclustering method binarizes the data matrix and locate submatrices with zero entries in all the cells. Spectral biclustering [79] uses eigen vectors to detect checkerboard structures in the data matrix that correspond to biclusters. SAMBA [80] formulates biclustering problem from graph theoretic view point and tries to find heavy subgraphs in a weighted bipartite graph. FLOC [81] starts with some initial seeds and then iteratively moves rows and columns to improve quality of biclustering with respect to a criterion function. ISA [82] is another biclustering method that incorporates randomization and finds biclusters with objects that possess constant values or coherently increasing values over the associated feature subset. To address the needs of biclustering problem effort is still underway [83], [84], [85], [86]

In comparison to the number of clustering and biclustering methods, there are not many triclustering methods. Jiang et al. [87] propose a set enumeration-based method to mine triclusters from 3-dimensional datasets using Pearson correlation coefficient. Authors propose two variant of the method to extract triclusters that are spread over all the time points. TRICLUSTER [88] is another well known triclustering method that extracts maximal triclusters from 3-dimensional datasets using a graph theoretic approach. gTRICLUSTER [89] uses Spearmen correlation coefficient to measure correlation among objects across time points while mining triclusters. ICSM [63] is a triclustering method that operates on possible pairs of time planes and detects some initial modules which are further extended to triclusters.

*3.3.2 Clustering methods for Big Data*

Though researchers have been trying to design clustering methods to address the issues mentioned above, none of these methods are seemed to handle all these issues simultaneously. Parallel clustering methods are seemed to be a solution for huge volume of data and incremental clustering methods handle high velocity data. Similarly, multi-view clustering methods are designed to handle data with variety.

The DBSCAN, DENCLUE, CLARA, CLARANS, and CURE methods discussed earlier are designed to handle large scale data. Two versions of k-means, namely k-mode and k-prototype methods [90] operate on large scale categorical and mixed type data, respectively. These methods use a low-complexity dissimilarity measure and cost function to make k-means suitable for large scale datasets. Ordonez et al. propose a variant of k-means [91] to minimize memory requirement and number of scans over the dataset. Bradley et al. propose a framework [92] to iteratively perform sampling from a large scale dataset and in each iteration, a model is improved to finally produce clusters. WaveCluster [93] uses wavelet transform to convert spatial domain data to frequency domain using a kernel function.

Li et al. propose parallel partitional and parallel single likange hierarchical clustering methods [94] on SIMD computers. Zhao et al. propose a parallel k-means clustering method that uses MapReduce architecture [95] to analyze parallel portions of the method. Similarly, PDBCSCAN [96] finds clusters from data distributed over multiple machines and the results are merged. P-cluster [97] partitions the objects to minimize the error. PBIRCH [98] is a parallel version of BIRCH that continuously distributes the incoming data among multiple processors using message passing and shared nothing architecture.

There exists incremental clustering methods to accomodate new objects without rerunning the clustering method on old objects. Chakraborty et al. propose an incremental k-means clustering [99] that computes new cluster centers by only using the existing cluster centers and the newly arrived object. Widyantoro et al. propose an incremental hierarchical clustering [100] that restructures the region of the object hierarchy by inserting the new object through a sequence of restructuring processes. IGDCA [101] is an incremental density based clustering method that divides data space into units with high density to form clusters. Insertion or deletion of an object only affects density of the unit to which it belongs.

Though multi-view learning is mostly in supervised and semi supervised learning, there are a few works that addresses the problem of unsupervised multi-view learning. Kailing et al. propose a multi-view clustering method [102] based on DBSCAN clustering method. The method operates on different feature spaces of various objects separately without combining these spaces. Zeng et al. propose a framework [103] that performs clustering in different feature spaces separately and then iteratively project and propagate the clustering results in multiple graph-based link layers until they converge. Chaudhuri et al. propose a multi-view clustering method [104] to project multi-views to lower dimensional space. The method tries to locate low dimensional subspace using a subspace learning method based on canonical correlation analysis. Kumar et al. propose a multi-view version of spectral clustering [105] that computes eigen vectors in different feature spaces and use these eigen vectors to improve graph structures of other view iteratively. The authors propose another multi-view version of spectral clustering [105] method that uses coregulation-based approaches to find coherent eigen vectors from different graphs.

## 3.4 Deep Learning

Deep learning attempts to model high-level abstractions in data using supervised and/or unsupervised learning algorithms, in order to learn from multiple levels of abstractions. It uses hierarchical representations of data for classification. Deep learning methods have been used in many applications, viz., pattern recognition, computer vision, natural language processing and speech recognition. Due to exponential increase of data in these applications, deep learning is useful for accurate prediction from voluminous data. In recent years, reseachers have developed effective and scalable parallel algorithms for training deep models [106]. Many organizations use deep learning for decision making, information retrieval, and semantic indexing. A deep learning architecture is shown in Figure 7. Input data are partitioned into multiple samples for data abstarctions. The intermediate layers are used to process the features at multiple levels for prediction from data. The final prediction is performed at the output layer using the outputs of its immediate upper layer.

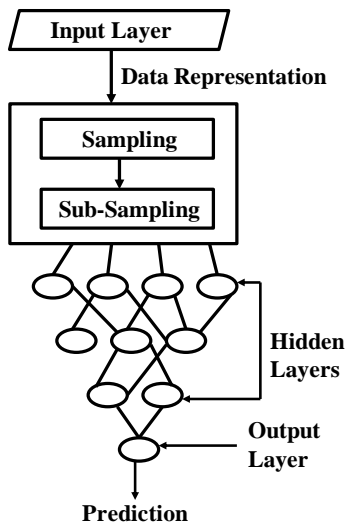

Fig. 7. Deep Leaning Architecture

Deep learning represents data in multiple layers. It can efficiently process high volume of data, where shallow learning fails to explore due to the complexities of data patterns. Moreover, deep learning is quiet suitable for analyzing unstructured and heterogeneous data collected form various sources.

Traditional neural networks pose two problems, viz., poor performance due to local optima of a non-convex object function and incapability to exploit unlabeled data, which are abundant and cheap. To overcome these limitations of traditional neural networks, Deep Belief Networks (DBN) [107] was introduced with a deep learning architecture to learn from both labeled and unlabeled data. The deep architecture of DBN integrates unsupervised pre-training and supervised fine-tuning strategies. The unsupervised pre-training is used to learn data distribution, whereas the supervised fine-tuning is used for local optima search. Ngiam et al. [108] propose a deep learning method by integrating both audio and video data for learning representations. The solution is effective in learning from multiple abstractions and can capture correlations across multiple abstractions.

Big data are continuously generated at a very high speed and require fast processing. Therefore, learning solution should not only be fast and efficient, but also be able to handle incremental data. However, we could not find a deep learning method that considers incremental data. Moreover, deep learning can handle only volume and variety of big data [109].

### 3.5 Inference of Large Scale GRN with Association Rule Mining

In system biology, complex dynamic behavior of a group of genes and how it influences the expression of other genes, may be represented as GRN. By comparing between normal and diseased networks, one can identify potential drug targets for the target disease [110]. Research labs are producing a large number of expression data and consequently, the state-of-the-art inference techniques are insufficient in handling such large scale GRN. Microarray experiments conducted in different growth environments leads to heterogeneous data. Presence of steady-state and perturb expression data make the task of inference more challenging. Looking into the magnitude of difficulties in handling such voluminous, continuous, and heterogeneous data, the task of GRN inference may be considered as a big data analytics problem [111], [112]. Specialized inference methods in big data paradigm are very much necessary. A number of inference methods have been proposed for last several years [113], [114], [115], [116], [117], [118], [119]. However, they are limited in handling data sets with more than thousands of genes. In most of the cases, execution performance degrades exponentially with the increase in number of nodes or genes in the network. The scenario becomes more adverse with the increase in number of conditions (dimensions) or time points in time series expression data.

#### 3.5.1 Serial Association Mining

Association rule mining (ARM) came into existence as market basket analysis on boolean datasets. In association mining the sizes of datasets are semi large that can usually be accommodated on main memory. Typically, they are static in nature.

The AIS (Agrawal, Imielinski, Swami) [120] and Apriori [121] are two pioneering algorithms for mining association rule Although, they are robust, two major limitations are that they generate too many candidate itemsets and require too many passes over the whole database. SETM [122] was motivated by the desire to use SQL to calculate large itemsets, whereas, DHP (Direct Hashing and Pruning) attempts to reduce the number of candidate itemsets [123].

The partition approach [124] mines frequent itemsets from large datasets by dividing into smaller partitions, whereas, sampling [125] reduces the number of database scans. DIC (Dynamic Itemset Counting) [126] drastically reduces the number of scans of the database during frequent itemset finding. FP-Growth [127] finds frequent itemsets without candidate generation. However, the time taken to construct the FP-tree is quite large and its performance degrades with the increase in support count. Recently, another effective algorithm called OPAM [128] has been proposed, for finding all the frequent itemsets without generating any candidate sets. OPAM adopts an integrated approach to solve the frequent itemset finding problem in a single pass over the database.

Today, most real-world databases are heterogeneous in nature, contain only quantitative data or both quantitative and categorical data. Further, such databases are multi-dimensional and their volume seems to be large. And this is where the conventional ARM techniques almost fail to satisfy the demand of mining fast growing varied voluminous data.

Attribute partitioning approach is the most evident one to deal quantitative attributes [129]. As reported in [130], a possible solution to figure out meaningful quantitative regions for the discovery of association rules is clustering approach.

### 3.5.2 Distributed and Parallel Association Mining

Sequential techniques are inadequate to provide scalability in terms of dimension, size or data which are spread around geographically dispersed locations. To cope up with such circumstances, researchers are looking for high-performance parallel and distributed association mining techniques. In the yester-years, a number of such techniques have been developed. These are mostly the extensions of already existing sequential methods.

The Count Distribution [131] is a simple parallelization of Apriori. This algorithm minimizes communication, because only the counts are exchanged among the processors. However, the algorithm replicates the entire hash tree on each processor and does not use the aggregate system memory effectively. PDM [132] is based on DHP [123]. In PDM, each processor generates the local supports of 1-itemsets and approximate counts for the 2-itemsets with a hash table. Next, PDM obtains the local counts for all candidates and exchanges them among all processors to determine the globally frequent itemsets. FDM (Fast Distributed Mining) [133] builds on Count Distribution [131], and the authors propose new techniques to reduce the number of candidates considered for counting and hence minimizes communication. FDM also suggests three optimizations: local pruning, global pruning, and count polling. To address the issues of FDM, a parallel version of FDM, called Fast Parallel Mining (FPM) [134] was introduced. FPM generates fewer candidates and retains the local and global pruning steps. But instead of count polling and subsequent broadcast of frequent itemsets, it simply broadcasts local supports to all processors.

Other than improvement in computational cost, distribute or parallel versions inherently carry all the demerits suffered by their respective serial methods. Recently, several efforts have been made to extend some of the serial rule mining methods to be implemented in MapReduce framework for faster execution [135] and handling for voluminous data.

### 3.5.3 Dynamic Association Mining

The techniques discussed in the previous sections are mostly based on the assumption that the datasets used as input does not change. In practice, no transaction database is static. Subsequent update of dataset could potentially invalidate existing association rules. Database updates require rediscovering the rules afresh by scanning the entire old and new data. Rediscovering of rules with the updates of the database leads to time consuming computation and leads to significant I/O overheads. The dynamics of databases can be represented as i) incremental updates and ii) decremental updates. A number of efficient techniques have been developed for mining dynamic datasets.

Fast UPdate (FUP) [136] was proposed to compute large itemsets in a dataset that is updated regularly. The framework of FUP is similar to that of Apriori and DHP and is referred to as a k-pass algorithm because it scans the dataset k times. The Borders [137] algorithm is based on the concept of border sets introduced in [138]. It is another incremental method to generate frequent sets. The Decrement Updating Algorithm [139] tries to detect all the frequent itemsets from dynamically deleted databases.

For faster handling of varied, voluminous data, current association mining techniques are inadequate. Big data paradigm demands an integrated solution encompassing almost all the approaches to handle dynamic, large, and heterogeneous data. Several attempts have been made to infer GRN based on steady-state time series data. However, none of them can handle dynamic time-series data [112]. There is an urgent need of a scalable GRN reconstruction method that can work to infer reliable GRNs.

## 4 CHALLENGES AND ISSUES IN BIG DATA ANALYTICS

Bioinformatics research has rapidly become a big data problem in the recent years. Big data not only possesses volume, velocity, and variety, but also are incremental and distributed. These properties of big data make it extremely difficult for the traditional data analytics to perform fast and accurately. Machine learning methods may be useful in handling big data analytics, since they have evolved in the computer science domain with objectives like performance and efficiency. The machine learning techniques for bioinformatics, the existing ones as well as those developed for handling big data, are discussed in the previous section. This section summarizes some of the challenges and research issues in big data analytics using machine learning methods.

### 4.1 Challenges in big data analytics

The techniques used for analysis and visualization of traditional databases are not adequate on big data. The volume, velocity, variety, distributedness, and incremental nature of such data impose challenges on the traditional methods for data analytics. The volume of data generation and the speed of data transmission are growing rapidly. Napatech, a manufacturer of high speed network accelerators reported in 2014 that all network data will grow with an annual growth rate of 23% through 2018. The exponential increase in the use of hand-held devices and their associated sensors have mostly contributed to the growth of big data in the recent years.

Along with the increase in the data volume, the speed of data generation and transmission are also increasing. According to the Cisco report [140], the average mobile network connection speed in 2014 was 1,683 kbps, which will reach approximately 4.0 Mbps by 2019. Real time analytics on big data become more difficult with high data velocity. Although batch mode analytics may be scalable to high data velocity using distributed and parallel computing techniques, the slow I/O operations severely affect the analytics performance. In this era, I/O speed is lagging far behind computing speed, acting as the limiting factor of computational throughput.

Moreover, these continuously generated data are highly heterogeneous in nature. Traditional databases are arranged in terms of a set of defined schemas. Data warehouses store and update data following the extraction-transformation-loading operations. Since big data systems continuously fetch new data in high velocity and high variety from heterogeneous sources, a structured database, such as data warehouse, is not at all suitable for dynamic storage and real time retrieval.

Given these challenges, the traditional data analytics techniques, such as machine learning and statistical analysis, are inefficient with big data in their original form. Consequently, the problem of machine learning enabled analytics has to be studied from the perspective of big data.

Data privacy is another major challenge of big data analytics, particularly in the bioinformatics and healthcare domain. In order to protect sensitive information, data sources might use data anonymity or publish only partial data. Analytics on partial or anonymous data might be more complex and inefficient.

### 4.2 Issues in big data Analytics

Big data analytics require processing of massive amount of structured, semi-structured, poly-structured, and unstructured data, that grow over time. Real time analytics impose an additional requirement of time bound computation. Techniques from AI may be applied to find patterns and relations in unstructured data. Similarly, big data analytics can be scaled using parallel and distributed computing technologies, without compromising on accuracy of results. However, traditional data analytics on big data have certain issues regarding scalability and performance, which are discussed below.

1) An integrated big data analytics architecture that is fault tolerant and able to handle voluminous and varied data in batches as well as in a continuous stream in real time is still missing.
2) Distributed computing is the prime solution to handle the massive volume of big data. However, most of the AI, data mining, and statistical analysis approaches are not originally designed for distributed computation. Although distributed algorithms have been proposed in the literature [141], [142], [143], they are mostly academic research and lack robust implementation, considering various MapReduce frameworks.
3) A big data store does not have a uniform data format. Rather big data analytics need to process heterogeneous data captured through sensors of various types. Therefore, intelligent algorithms are required to find a coherent meaning from disparate data. This increases the complexity of analytics.
4) Unstructured, semi-structured and poly-structured data introduce more problems, such as data inconsistency and redundancy. Data pre-processing is costly due to their heterogeneous nature and massive volume. Traditional data analytics techniques attempting to handle inconsistent and noisy data are found costly in terms of time and space complexities.
5) Big data analytics need to mine datasets at different levels of abstraction. This significantly increases the complexity of the analytics methods, however, enabling biologists to analyze the data at various level of abstractions help understanding the interest of sematics biological data.

## 5 TOOLS FOR BIG DATA ANALYTICS IN BIOINFORMATICS

Various tools have been developed over the years to handle the bioinformatics problems. The tools developed before the big data era are mostly standalone and not designed for very large scale data. In the last decade many large scale data analysis tools have been developed for several problems, such as microarray data analysis to idetify coexpressed patterns, gene-gene network analysis and salient module extraction, PPI complex finding, and RNA/DNA and sequence analysis. However, apart from certain sequence analysis tools, the other exsting tools are not adequate for handling big data or not suitable for cloud computing infrastructures. Along with specific tools, several cloud-based bioinformatics platforms have also been developed to integrate specific tools and to provide a fast comprehensive solution to multiple problems, such as Galaxy [144] and CloudBLAST [145].

1) *Tools for microarray data analysis*
   Large number of software tools are available to perform various analysis on microarray data. However, not all the software are designed to handle large scale data. With the increase in the size of data sets, the time required to generate samples and sequences to identify complexes and to process heterogeneous disease query to find relevant complexes has become prohibitive. Beeline[22] handles big data size by parallel computations and reduction in the data size with adaptive filtering. A quality assurance tool called caCORRECT [146] removes artifactual noises from high throughput microarray data. caCORRECT may be used to improve integrity and quality of both public microarray archives as well as reproduced data and to provide a universal quality score for validation. A web-based application called omniBiomarker [147] uses knowledge-driven algorithms to find differentially expressed genes for biomarker identification from high throughput gene expression data. The approach requires complex computation and validation, and omniBiomarker helps in identifying stable and reproducible biomarkers.

2) *Tools for gene-gene network analysis*
   Gene expression datasets are already massive in size and getting bigger everyday. FastGCN [148] tool exploits parallelism with GPU architectures to find the co-expression networks in an optimized way. Similar GPU accelerated co-expression networks analysis methods are proposed by Arefin et al. [149] and McArt et al. [150]. The UCLA Gene Expression

---

22. illumina.com/applications/microarrays/microarray-software/beeline.html

Tool (UGET) [151] performs large scale co-expression analysis to find disease gene associations. Disease networks have significantly higher gene correlations and UGET calculates the correlations among all possible pairs of genes. UGET has been found effective when tested on Celsius [152], which is the largest co-normalized microarray dataset of Affymetrix-based gene expression datawarehouse. WGCNA [153] is a popular R package for performing weighted gene co-expression network analysis and can be used in an R-Hadoop distributed computing system.

3) *Tools for PPI data analysis*

PPI complex finding problem is a highly time consuming process. From our research experience, standalone implementations for both supervised and unsupervised PPI complex finding, such as MATLAB programs, require days or even weeks of time to find complexes from a dataset of approximately 1 million interactions on standard workstations. Therefore, there is an urgent need to develop fast big data tools for PPI complex finding and ranking, w.r.t. any heterogenious disease network query. Several relatively fast tools have been developed for PPI complex (isolated and overlapping) finding, such as NeMo [154], MCODE [155], and ClusterONE [156], either as a standalone tool or as a Cytoscape plugin. However, these tools cannot be used in distributed systems for better efficiency on large-scale PPI data. Finally, PathBLAST [157] is an important web-based tool for fast alignment of protein interaction networks.

4) *Tools for sequence analysis*

For sequence analysis problems, several tools have been developed on top of the Hadoop MapReduce platform to perform analytics on large scale sequence data. BioPig [158] is a notable hadoop-based tool for sequence analysis that scales automatically with the data size and can be ported directly to many hadoop infrastructures. SeqPig [159] is another such tool. The Crossbow [160] tool combines Bowtie [161], an ultrafast and memory efficient short read aligner, and SoapSNP [162], an accurate genotyper, to perform large scale whole genome sequence analytics on cloud platforms or on a local hadoop cluster. Other cloud-based tools that have been developed for large scale sequence analysis are Stormbow [163], CloVR [164], and Rainbow [165]. There exist other programs for large scale sequence analysis that do not use big data technologies, such as Vmatch [166] and SeqMonk[23].

5) *Tools for pathway analysis*

To support pathway analysis, a good number of tools have been developed for pathway analysis, such as GO-Elite [167] to describe particular genes or metabolites, PathVisio [168] for analysis and drawing, directPA [169] to perform analysis in a high-dimensional space for identifying pathways, Pathway Processor [170] to analyze expression data regarding metabolic pathways, Pathway-PDT [171] to perform analysis using raw genotypes in general nuclear families, and Pathview [172] for pathway based data integration. However, these tools neither use distributed computing platforms, nor they are developed as a cloud-based application, for high scalability.

---

23. www.bioinformatics.bbsrc.ac.uk/projects/seqmonk

Although evolutionary research may use tools developed for more specific problems, such as sequence analysis or gene-gene network analysis, a big data tool for comprehensive evolutionary research is still not known. The existing tools for evolution research, such as MEGA [173] and EvoPipes.net [174] are not developed for big data in evolutionary research.

# 6 CONCLUSION

This paper discusses the recent surge in bioinformatics data stores in terms of volume as well as dimension. With the advent of new high throughput and cheap data capturing tools, this rapid growth in data will continue in the coming years. Bioinformatics data are voluminous, heterogeneous, incremental, and distributed geographically all over the world. Consequently, the big data analytics techniques are required to solve the problems in bioinformatics.

The problems, data sources and data types in bioinformatics are diverse in nature. The existing big data architectures do not provide a comprehensive solution for big data analytics, which is fast, fault tolerant, large scale, incremental, distributed, and optimized for iterative and complex computations. The well known MapReduce architecture for distributed computing executes in a batch mode and has high disk read/write overhead. On the other hand, the graph-based architectures for streaming applications fail to provide fault tolerance. An integrated big data analytics architecture that fulfills the requirements of the problems in bioinformatics is an urgent need.

Machine learning has been the most utilized tool for data analytics. Large scale data existed well before the big data era, particularly in bioinformatics. Machine learning tools have been successfully used to analyze both small scale as well as large scale data using various techniques such as, sampling, feature selection, and distributed computations. However, big data poses more challenges on the traditional learning methods in terms of velocity, variety, and incremental data. Traditional learning methods usually embed iterative processing and complex data dependency among operations. Consequently, the traditional machine learning methods cannot be used to perform fast processing on massive data using big data platforms, such as MapReduce. This paper discusses the traditional machine learning methods , their limitations, and the efforts made in the recent years to extend them for big data, such as the incremental, parallel, and multi-view clustering methods to handle complex bioinformatics problems.

The existing tools for many bioinformatics problems are still not adequate for big data. A few tools have been developed for sequence analysis using the Hadoop MapReduce platform, such as BioPig [158] and Crossbow [160] in the recent years. Apart from that, other important bioinformatics problems, such as PPI network analysis or disease network analysis, still lacking Hadoop or cloud-based big data tools. Considering the big data boom in bioinformatics and the emerging research opportunities, big data analytics in bioinformatics need to be properly addressed from the perspectives of big data technologies and effective data analytics approaches, such as machine learning.

## ACKNOWLEDGMENTS

The authors would like to thank the Ministry of HRD, Govt. of India for funding as a Centre of Excellence with thrust area in Machine Learning Research and Big Data Analytics for the period 2014-2019.

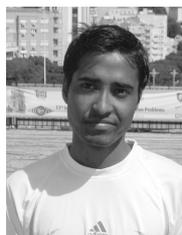

**Hirak Kashyap** obtained Master of Technology degree in Computer Science and Engineering from National Institute of Technology, Rourkela, India. Currently, he is a Senior Research Fellow in the Department of Computer Science and Engineering at Tezpur University. He conducts research in reconfigurable and embedded systems, architecture, and network security.

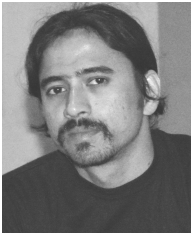
**Hasin A Ahmed** obtained Master of Computer Applications degree from Tezpur University, India in the year 2009. Currently, he is a PhD candidate in the Department of Computer Science and Engineering at Tezpur University. He conducts research in bioinformatics and data mining and has published his works in reputed journals.

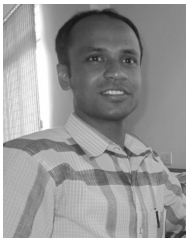
**Nazrul Hoque** obtained Master of Technology degree in Information Technology from Tezpur University, India in the year 2012. Currently, he is a PhD candidate in the Department of Computer Science and Engineering at Tezpur University. His research interests are machine learning and network security.

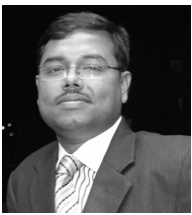
**Swarup Roy** obtained PhD degree in Computer Science and Engineering from Tezpur University, India in the year 2013. Currently, he is an Assistant Professor in Information Technology at North-Eastern Hill University, Shillong, India. His research interests are data mining and computational biology.

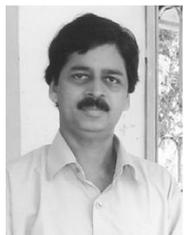
**Dhruba Kr Bhattacharyya** received his Ph.D. in Computer Science from Tezpur University in 1999. He is a Professor in the Computer Science & Engineering Department at Tezpur University. His research areas include data mining, bioinformatics, network security, and big data analytics. Prof. Bhattacharyya has published 220+ research papers in the leading international journals and conference proceedings. In addition, Dr Bhattacharyya has written/edited 8 books. He is a Programme Committee/Advisory Body member of several international conferences/workshops.